\documentclass[onecolumn,floatfix,aps,nofootinbib]{revtex4}

\usepackage[dvips]{epsfig}
\usepackage[english]{babel}
\usepackage{bbm}
\usepackage{verbatim}
\usepackage{array}
\usepackage{bm} 
\usepackage{amsmath}
\usepackage{amsfonts}
\usepackage{amssymb}
\usepackage{hyperref}
\usepackage[dvipsnames]{xcolor}
\hypersetup{colorlinks=true,linkbordercolor=Blue,linkcolor=Blue, citecolor=Blue}
\usepackage{indentfirst} 
\usepackage{float}

\usepackage{bbold}
\def\I{\openone}
\def\openone{\mathbb I}

\begin{document}

\title{Regular spinors and fermionic fields}

\author{R. J. Bueno Rogerio$^{1}$} \email{rodolforogerio@gmail.com}
\author{J. M. Hoff da Silva$^{2}$} \email{julio.hoff@unesp.br}
\author{C. H. Coronado Villalobos$^{3}$} \email{ccoronado@icalidad.gob.pe}
\affiliation{$^{1}$Instituto de F\'isica e Qu\'imica, Universidade Federal de Itajub\'a - IFQ/UNIFEI, \\
Av. BPS 1303, CEP 37500-903, Itajub\'a - MG, Brazil.}
\affiliation{$^{2}$Departamento de F\'isica, Universidade
Estadual Paulista, UNESP, Av. Dr. Ariberto Pereira da Cunha, 333, Guaratinguet\'a, SP,
Brazil}
\affiliation{$^{3}$Instituto Nacional de Calidad - INACAL, Calle las Camelias, 817, San Isidro, Lima, Per\'u.}


\begin{abstract}
Bearing in mind the Lounesto spinor classification, we connect the expansion coefficients of well behaved fermionic quantum field, i.e., a local field within a full Lorentz covariant theory, with and only with a given subclass of Type-2 spinors according to Lounesto. We comment on theoretical possibilities as well as physical outputs for the other cases. 
\begin{center}
	\emph{In memory of Allan Yoffre Coronado Villalobos.}
\end{center} 
\end{abstract}


\maketitle

\section{Introduction}\label{intro}

Roughly speaking, the foundations of Quantum Field Theory dictates that quantum fields are the result of engaging well defined one particle states \cite{Wigner1} into interactions regulated, so to speak, by Lorentz covariance \cite{weinbergfeynman}. Besides, fundamental rules as the respect to the cluster decomposition principle \cite{wic} and (micro)causality, in the sense of Weinberg, do form the theoretical scope upon which a quantum field arises \cite{weinberg1}. 

Concerning spin one half particles, and corresponding fields, the aforementioned formulation has gained the status of a (no-go) theorem. Even never been enunciated as such initially, the formulation is exhaustive enough, however, for such an epithet. It asserts that a local and Lorentz invariant quantum field whose action in a vacuum state leads to spin one half particles is an usual Dirac field. There are two utterly relevant points in the whole formulation whose appreciation has been proved quite pertinent: the fermionic dual structure and the role of the parity operator. The (necessary) theory for the definition of a given adjoint structure in fermionic theory can be found in Ref. \cite{aaca,mdobook}, and its departure from the usual Dirac case for mass dimension one fermions was explored to circumvent the Weinberg no-go theorem \cite{nogo}, while for the very same field the role of parity was further explored evincing the one particle states of such field as belonging to an specific non standard Wigner class \cite{elkostates}. In this work we shall further explore the relation of quantum fermionic fields and the use of parity symmetry in their construction. More specifically we relate the fermions satisfying the Dirac equation - those entering in the no-go theorem - with a subset of a given particular class of spinors appearing in a classical classification, according to its bilinear covariants, due to Lounesto \cite{lounestolivro}. It is also shown that spinors not belonging to such a subset do not satisfy Dirac dynamics and lead to a non local quantum field (at least when the field adjoint is the usual one). Besides, their quantum states are unusual in a precise sense, we made it clear along the text. 

The Lounesto classification makes use of the Fierz-Pauli-Kofink identities and the inversion theorem \cite{tak} (allowing for expressing spinors in term of their bilinear covariants) to categorize spinors into six disjoint classes (see Ref. \cite{out} for a short review). Three of these classes, with non vanishing scalar ($\bar{\psi}\psi$) and/or pseudo-scalar ($\bar{\psi}\gamma_5\psi$) are called regular spinors. We shall deal in this paper with these regular spinors. The other sector, composed by singular spinors, have always null orthonormality relation and the standard physical interpretation may be obscured. 

We show that using a general spinor playing the role of expansion coefficient function of a fermionic quantum filed, the imposition of Dirac dynamics (at classical level) leads to a local quantum field within a theory respecting Lorentz symmetries and the coefficients are automatically restricted to a subclass, say $L_2$, of Type-2 spinors, according to Lounesto. For regular spinors of Type-2 other than the spinors belonging to $L_2$ subclass, and spinors belonging to other types, locality is not directly ensured and Dirac dynamics is not fulfilled. This approach may be seen as a link between the Weinberg no-go theorem and the Lounesto classification, evincing that even having a plethora of possibilities coming from the classical analysis, Dirac dynamics (or full Lorentz invariance) associated with the demand for locality restricts the possibilities of fermionic quantum fields (for which the dual is the Dirac one). With respect to this fact, we comment on the new possibilities coming from the fermionic dual theory \cite{mdobook} and speculate on the classification of its possible one particle states, as well.  

All the results of this work may be found in the next section, where we connect Type-2 spinors with a well behaved fermionic quantum field interpreting and discussing the results. In a subsection we develop an explicit example of non-local field and ended by addressing some comments on the possibilities which can arise in the scope of the fermionic dual theory. In the final section we conclude.

\section{Types and Phases}

We start by defining the eigenstates of the helicity operator, $\vec{\sigma}\cdot\hat{p}$, which reads
\begin{equation}\label{operadorhelicidade}
\vec{\sigma}\cdot\hat{p}\; \phi^{\pm}(k^{\mu}) = \pm \phi^{\pm}(k^{\mu}),
\end{equation}
where $\sigma$ stands for the Pauli matrices and the momentum unit vector reads $\hat{p}=(p\sin(\theta)\cos(\phi),\sin(\theta\sin(\phi), p\cos(\theta))$. We shall use these states as the spinors composing the bispinors representation. In the rest-frame referential the positive and negative helicity components read 
\begin{eqnarray}\label{components}
\phi^{+}(k^{\mu}) = \sqrt{m}\left(\begin{array}{c}
\cos(\theta/2)e^{-i\phi/2} \\ 
\sin(\theta/2)e^{i\phi/2}
\end{array}\right), \;\;  \phi^{-}(k^{\mu}) = \sqrt{m}\left(\begin{array}{c}
\sin(\theta/2)e^{-i\phi/2} \\ 
-\cos(\theta/2)e^{i\phi/2}
\end{array}\right). 
\end{eqnarray} 

A careful remark presented in Ref \cite{aaca,mdobook} elucidates the importance of additional relative phase factors in \eqref{components}. Such factors indeed play an important role in the study of the spinors behavior under discrete symmetries and may also be used to ensure locality for the quantum fields defined upon those expansion coefficients. 

Here, as we shall see in brief, we introduce, as a strategy, (disguised) relative trial phase factors in order to relate the behavior of spinors under parity (and explore locality issues) with its position, so to speak, in the Lounesto classes classification. In this regard, we shall not fix the phases \emph{a priori}, but instead to determine it relating the appropriate behaviors with the Lounesto classes in first and second quantization context. We opted for introducing two trial phases for the sake of exposition and interpretation. 

Let $\psi$ to be a single-helicity spinor in Weyl representation
 \begin{eqnarray}
\psi_{(+,+)}(k^{\mu}) = \left(\begin{array}{c}
e^{i\alpha}\phi^{+}(k^{\mu})\\
e^{i\beta}\phi^{+}(k^{\mu})
\end{array}
\right), \;\; \psi_{(-,-)}(k^{\mu}) =  \left(\begin{array}{c}
e^{i\alpha}\phi^{-}(k^{\mu})\\
e^{i\beta}\phi^{-}(k^{\mu})
\end{array}
\right),
\end{eqnarray}
in which $\alpha\in{\rm I\!R}$ and $\beta\in{\rm I\!R}$, standing for the alluded phases. To define the spinors for an arbitrary momentum referential, we define $\psi(p^{\mu})=\kappa\psi(k^{\mu})$, where the Lorentz boost operator is given by
\begin{eqnarray}
\kappa = \sqrt{\frac{E+m}{2m}}\left(\begin{array}{cc}
\I+ \frac{\vec{\sigma}\cdot\vec{p}}{E+m} & 0 \\ 
0 & \I- \frac{\vec{\sigma}\cdot\vec{p}}{E+m}
\end{array} \right).
\end{eqnarray} 
Calling $\mathfrak{B}_{\pm} = \sqrt{\frac{E+m}{2m}}\big(1\pm\frac{p}{E+m}\big)$ the appropriated Lorentz boosts factors, we may define a set of single-helicity spinors in the following fashion
\begin{equation}\label{psisingleparticula}
\psi^{\mathtt{P}}_{(+,+)}(p^{\mu}) = \sqrt{m}\left(\begin{array}{c}
e^{i\alpha}\mathfrak{B}_{+}\cos(\theta/2)e^{-i\phi/2} \\ 
e^{i\alpha}\mathfrak{B}_{+}\sin(\theta/2)e^{i\phi/2} \\ 
e^{i\beta}\mathfrak{B}_{-}\cos(\theta/2)e^{-i\phi/2} \\ 
e^{i\beta}\mathfrak{B}_{-}\sin(\theta/2)e^{i\phi/2}
\end{array}\right),\; \psi^{\mathtt{P}}_{(-,-)}(p^{\mu}) = \sqrt{m}\left(\begin{array}{c}
-e^{i\alpha}\mathfrak{B}_{-}\sin(\theta/2)e^{-i\phi/2} \\ 
e^{i\alpha}\mathfrak{B}_{-}\cos(\theta/2)e^{i\phi/2} \\ 
-e^{i\beta}\mathfrak{B}_{+}\sin(\theta/2)e^{-i\phi/2} \\ 
e^{i\beta}\mathfrak{B}_{+}\cos(\theta/2)e^{i\phi/2}
\end{array}\right).
\end{equation} Moreover, taking into account the judicious constraint: $\phi_{R}(k^{\mu})=-\phi_{L}(k^{\mu})$, see a complete discussion in Refs \cite{ahluwalia1ryder,gaioliryder}, we may also define the following spinors related to anti-particles
 \begin{equation}\label{psisingleantiparticula}
\psi^{\mathtt{A}}_{(+,+)}(p^{\mu}) = \sqrt{m}\left(\begin{array}{c}
-e^{i\alpha}\mathfrak{B}_{+}\cos(\theta/2)e^{-i\phi/2} \\ 
-e^{i\alpha}\mathfrak{B}_{+}\sin(\theta/2)e^{i\phi/2} \\ 
e^{i\beta}\mathfrak{B}_{-}\cos(\theta/2)e^{-i\phi/2} \\ 
e^{i\beta}\mathfrak{B}_{-}\sin(\theta/2)e^{i\phi/2}
\end{array}\right),\; \psi^{\mathtt{A}}_{(-,-)}(p^{\mu}) = \sqrt{m}\left(\begin{array}{c}
e^{i\alpha}\mathfrak{B}_{-}\sin(\theta/2)e^{-i\phi/2} \\ 
-e^{i\alpha}\mathfrak{B}_{-}\cos(\theta/2)e^{i\phi/2} \\ 
-e^{i\beta}\mathfrak{B}_{+}\sin(\theta/2)e^{-i\phi/2} \\ 
e^{i\beta}\mathfrak{B}_{+}\cos(\theta/2)e^{i\phi/2}
\end{array}\right).
\end{equation}
Of course the indexes $\mathtt{P}$ and $\mathtt{A}$ stand for particle and anti-particle, respectively. It is important to remark that since we have fixed the basis the solutions with $\alpha$ and $\beta$ are not a unitary transformation of the standard Dirac spinors.


It is a good point to recall the set-up underling the Lounesto spinor classification \cite{lounestolivro}. For sections of $P_{SL(2,\mathbb{C})}\times_\rho \mathbb{C}^4$, where $\rho=(1/2,0)$, $(0,1/2)$, or $(1/2,0)\oplus(0,1/2)$, which is indeed the case for the spinors just outlined, Lounesto shown the existence of six dense and disjoint classes of spinors based in the behavior of the their bilinear covariants. The basic strategy for the classification is the use of the so-called inversion theorem \cite{tak}, by means of which it is possible to express the spinor in terms of its own bilinear covariants and constraint the several possibilities via Fierz-Pauli-Kofink identities \cite{out}. From among these classes, three of them have in common the fact that the scalar and/or pseudo-scalar bilinear are non-null. These are the regular classes of spinors and they are organized as follows: Type-1 spinors have both, scalar and pseudo-scalar, bilinear covariants non-vanishing. Type-2 spinors have the scalar non-null and vanishing pseudo-scalar while the opposite is true for Type-3 spinors. If these two bilinear covariants are null the classification gives another classes, all of them organized as singular spinors. 

Returning to the case at hands, we remark that without a judicious inspection of $\alpha$ and $\beta$ it is impossible to ascertain the class each spinor above belongs to. The analysis of the proper Lounesto classification of the general single-helicity spinors in terms of their phases leads to important cases and sub cases \cite{rodolfoconstraints} summarized in the Table I. We emphasize that any other relation is a subsidiary condition from one of the constraints displayed in Table I. 

\begin{table}
\centering
\begin{tabular}{c|c|c|c}
\hline 
 \;\;\;\;\;\;\;\;\;\;$\alpha=\mathfrak{n}\pi$\;\;\;\;\;\;\;\;\;\; & \;\;\;\;\;\;\;\;\;\;$\beta=\mathfrak{m}\pi$\;\;\;\;\;\;\;\;\;\; & \;\;\;\;\;\;\;\;\;\;Type\;\;\;\;\;\;\;\;\;\; & Constraints \\ 
\hline 
\hline 
 $\mathfrak{n}=non-integer$ & $\mathfrak{m}=non-integer$ & 1 & $\mathfrak{n}\neq k\mathfrak{m}$, with $k = integer$. \\ 
 $\mathfrak{n}=integer$ & $\mathfrak{m}=non-integer$ & 1 & $\mathfrak{m}\neq \mathfrak{n}/k$, with $k>2$. \\
  $\mathfrak{n}=non-integer$ & $\mathfrak{m}=non-integer$ & 1 & $\mathfrak{m}= 1/\mathfrak{n}$. \\
 $\mathfrak{n}=integer$ & $\mathfrak{m}=integer$ & 2 & $\mathfrak{n}=\mathfrak{m}$. \\ 
$\mathfrak{n}=integer$ & $\mathfrak{m}=integer$ & 2 & $\mathfrak{n}\neq \mathfrak{m}$.\\ 
$\mathfrak{n}=non-integer$ & $\mathfrak{m}=non-integer$ & 2 & $\mathfrak{n}= k\mathfrak{m}$, with $k = integer$. \\
  $\mathfrak{n}=non-integer$ & $\mathfrak{m}=integer$ & 3 &  $\mathfrak{n}=k/2$, with $k=odd$. \\
 \hline 
\hline 
\end{tabular}\label{tabela1}
\caption{The main phases constraints to classify regular spinors via their relative phases. The type is also preserved by interchanging the conditions for $\mathfrak{n}$ and $\mathfrak{m}$ \cite{rodolfoconstraints}.}
\end{table} 

As one can see the trial phases engender a mapping between regular spinors and the plane $(\alpha,\beta)\simeq \mathbb{R}^2$. Besides, as will be clear in the following, the subclass $L_2$ of Type-2 spinors for which $\alpha=\beta$ (or $\mathfrak{n}=\mathfrak{m}$) is special. For this subclass, and only for this subclass, (correspondingly for the straight line $\mathbb{R}^2|_{\alpha=\beta}\simeq \mathbb{R}\subset\mathbb{R}^2$) the Dirac dynamics is ensured for the classical spinors and locality is attained for the associated quantum field. 

In order to approach the aforementioned results, we start computing the spin sums. As a central aspect of any quantized field, the spin sums shall enter in the analysis of quantum correlators and, as expected, give an important clue to the field dynamics. Hence, the spin sums associated to (\ref{psisingleparticula}) and (\ref{psisingleantiparticula}) read, respectively 

\begin{eqnarray}\label{spinsumparticula}
\sum_{h}\psi^{\mathtt{P}}_{h}(p^{\mu})\bar{\psi}^{\mathtt{P}}_{h}(p^{\mu})=
\left(\begin{array}{cccc}
m e^{i(\alpha-\beta)}  & 0 & E+p\cos\theta & p\sin\theta e^{-i\phi} \\ 
0 & m e^{i(\alpha-\beta)} & p\sin\theta e^{i\phi} & E-p\cos\theta \\ 
E-p\cos\theta & -p\sin\theta e^{-i\phi} & m e^{-i(\alpha-\beta)}  & 0 \\ 
-p\sin\theta e^{i\phi} & E+p\cos\theta & 0 & m e^{-i(\alpha-\beta)} 
\end{array} \right),
\end{eqnarray}
and 
\begin{eqnarray}\label{spinsumanti}
\sum_{h}\psi^{\mathtt{A}}_{h}(p^{\mu})\bar{\psi}^{\mathtt{A}}_{h}(p^{\mu})=\left(\begin{array}{cccc}
-m e^{i(\alpha-\beta)} & 0 & E+p\cos\theta & p\sin\theta e^{-i\phi} \\ 
0 & -m e^{i(\alpha-\beta)} & p\sin\theta e^{i\phi} & E-p\cos\theta \\ 
E-p\cos\theta & -p\sin\theta e^{-i\phi} & -m e^{-i(\alpha-\beta)} & 0 \\ 
-p\sin\theta e^{i\phi} & E+p\cos\theta & 0 & -m e^{-i(\alpha-\beta)} 
\end{array} \right).
\end{eqnarray} It is possible to recast the result more appropriately, in a Dirac-like fashion, as 

\begin{eqnarray}
&&\sum_{h}\psi^{\mathtt{P}}_{h}(p^{\mu})\bar{\psi}^{\mathtt{P}}_{h}(p^{\mu})= \gamma_{\mu}p^{\mu} + m \I_{(\alpha,\beta)},
\\
&&\sum_{h}\psi^{\mathtt{A}}_{h}(p^{\mu})\bar{\psi}^{\mathtt{A}}_{h}(p^{\mu})=\gamma_{\mu}p^{\mu} - m \I_{(\alpha,\beta)}, 
\end{eqnarray} where $\gamma^\mu$ matrices are in the Weyl representation and the $\I_{(\alpha,\beta)}$ matrix stands for
\begin{eqnarray}
\I_{(\alpha,\beta)} = \left(\begin{array}{cccc}
e^{i(\alpha-\beta)}  & 0 & 0 & 0 \\ 
0 & e^{i(\alpha-\beta)}  & 0 & 0 \\ 
0 & 0 & e^{-i(\alpha-\beta)}  & 0 \\ 
0 & 0 & 0 & e^{-i(\alpha-\beta)} 
\end{array} \right).\label{tra}
\end{eqnarray}
Despite formal similarity with the usual Dirac equation, spinors which do not exactly fulfill Dirac dynamics $(\alpha=\beta)$ experience locality differently. We also call attention to the following fact: $\det(\gamma_{\mu}p^{\mu}\pm m\I)=0$ and $\det(\gamma_{\mu}p^{\mu}\pm m\I_{(\alpha,\beta)})=0$, both yielding the usual dispersion relation $E=\pm\sqrt{m^2+p^2}$.

It is then trivial to see that the spin sums are covariant under Lorentz symmetries and they indeed completeness relation
\begin{eqnarray}
\frac{1}{2m}\sum_{h}\big[\psi^{\mathtt{P}}_{h}(p^{\mu})\bar{\psi}^{\mathtt{P}}_{h}(p^{\mu})-\psi^{\mathtt{A}}_{h}(p^{\mu})\bar{\psi}^{\mathtt{A}}_{h}(p^{\mu})\big] =\I_{(\alpha,\beta)},
\end{eqnarray} as expected. Moreover for the special case performed by the subclass of Type-2 spinors for which $\alpha=\beta$ the spin sums are nothing but the usual textbook ones. The $\alpha=\beta$ imposition may also be appreciated from the acting of Dirac operator into the spinors at hand. We illustrate this claim by picking the spinor $\psi^{\mathtt{P}}_{(+,+)}(p^{\mu})$, for example. A straightforward calculation leads to 
\begin{equation}\label{diracnotype1}
(\gamma_{\mu}p^{\mu}-m)\psi^{\mathtt{P}}_{(+,+)}(p^{\mu}) = \sqrt{m}\left(\begin{array}{c}
m [e^{i\beta}-e^{i(2\alpha-\beta)}]\mathfrak{B}_{+}\cos(\theta/2)e^{-i\phi/2} \\ 
m [e^{i\beta}-e^{i(2\alpha-\beta)}]\mathfrak{B}_{+}\sin(\theta/2)e^{i\phi/2} \\ 
m [e^{i\alpha}-e^{i(2\beta-\alpha)}]\mathfrak{B}_{-}\cos(\theta/2)e^{-i\phi/2} \\ 
m [e^{i\alpha}-e^{i(2\beta-\alpha)}]\mathfrak{B}_{-}\sin(\theta/2)e^{i\phi/2}
\end{array}\right). 
\end{equation}
Therefore, the spinor is annihilated by the Dirac operator at the classical level if, and only if, $\alpha=\beta$ and we are lead to a subclass of Type-2 spinors. This is the alluded $L_2$ subclass. It may be verified that as far as we are concerned with this particular subclass then, indeed, $(\gamma_\mu p^\mu\mp m)\psi_h^{\mathtt{P/A}}(p^\mu)=0$. It was demonstrated in Ref \cite{speranca} that, at the classical level, Dirac dynamics is related to the parity operator, i. e., $P\psi(p^\mu)=m^{-1}\gamma^\mu p_\mu \psi(p^\mu)$ $\forall$ $\psi(p^\mu)\in P_{SL(2,\mathbb{C})}\times\mathbb{C}^4$. Therefore the requirement that a given spinor be eingenspinor of $P$ implies Dirac dynamics. Within this perspective, it is clear that different phases inserted in the Weyl spinors ($\alpha\neq\beta$) would lead to a violation in the Dirac dynamics: the different sides of the representation space (embracing left or right hand spinors) are not being equally treated. Hence parity is being violated at classical level and, consequently, the Dirac operator does not annihilate the resulting bi-spinor. We also notice that the matrix in (\ref{tra}) is related to $\gamma_5$ in a different disguise, i. e., $ \I_{\alpha,\beta}=\exp[i \gamma^5 (\alpha-\beta)]$.

The above observations may be followed by some additional characterization of the mapping between the $\mathbb{R}^2$ plane $(\alpha, \beta)$ and the regular spinors. The matrix (\ref{tra}) is a relevant output of the introduced trial phases. This matrix is characterized by the mapping 
\begin{eqnarray} 
\I_{(\alpha,\beta)}:\mathbb{R}^2\rightarrow M_{4\times 4}(\mathbb{C}) \nonumber \hspace{4.3cm} &&\\
(\alpha,\beta)\mapsto \left(\begin{array}{cc}
e^{i(\alpha-\beta)}  & 0  \\ 
0& e^{-i(\alpha-\beta)} \\  
\end{array} \right)\otimes  \left(\begin{array}{cc}
1 & 0  \\ 
0 & 1 \\  
\end{array} \right).
\end{eqnarray} Notice that the already pointed the sector of $\mathbb{R}^2$ corresponding to $L_2$ spinors (for which $\I_{(\alpha,\beta)}\simeq \I$), $\mathbb{R}^2|_{\alpha=\beta}\simeq \mathbb{R}$, is topologically trivial. However, the complement to $\mathbb{R}^2$, that is, the portion of the plane corresponding to Type-1, Type-3, and Type-2 spinors for which $\alpha\neq \beta$, is clearly given by $\mathbb{R}^2\backslash (\mathbb{R}^2|_{\alpha=\beta}\simeq \mathbb{R})$ which is not connected\footnote{Besides, by means of an artificial, but ludic, analogy we may extract more differences between the situation performed by spinors belonging to $L_2$ and their counterpart sector in the $(\alpha,\beta)$ plane. Bearing in mind the range of $\I_{(\alpha,\beta)}$ one is able to study its behavior under partial derivatives. It is fairly simple to see that $\partial_\alpha\I_{(\alpha,\beta)}=-\partial_\beta\I_{(\alpha,\beta)}$. In this regard, it may be useful to think of $\I_{(\alpha,\beta)}$ as components of a special  ``vector'' $\mathcal{V}=\I_{(\alpha,\beta)}\hat{\alpha}+\I_{(\alpha,\beta)}\hat{\beta}$ which, under acting of $\nabla:=\hat{\alpha}\partial_\alpha +\hat{\beta}\partial_\beta$, shows itself everywhere divergence-free in the plane $(\alpha,\beta)$
	\begin{equation}
	\nabla\cdot\mathcal{V}=(\partial_\alpha+\partial_\beta)\I_{(\alpha,\beta)}=0. \nonumber
	\end{equation} This last equation is always satisfied but it is trivialized for $\mathbb{R}^2_{\alpha=\beta}\simeq \mathbb{R}$. In this region, and only there, the field $\mathcal{V}$ is also (pseudo)irrotational and conservative, with potential scalar function given by the remain phase multiplying the identity matrix. We use the adjectivation ``pseudo'' here to account on the dimension at hands. Without any worry to this analogy one can formalize the rotational of $\mathcal{V}$ in an artificial $(\hat{\alpha}\times\hat{\beta})$-direction. For every pair $(\alpha,\beta)$ there is a corresponding regular spinor belonging to a given type, according to Lounesto. The special cases for which the ``vector'' field composed by $\I_{(\alpha,\beta)}$ is conservative is given by spinors belonging to $L_2$, respecting Dirac dynamics. }, since $\pi_0(\mathbb{R}^2\backslash \mathbb{R})\neq 0$.

Finally, returning to the physical aspects, it is possible to define quantum fields based in expansion coefficients performed by single-helicity spinors belonging to $L_2$ endowed with $\alpha=\beta$($=0$, for simplicity) 
\begin{eqnarray}\label{campoquantico}
\mathfrak{F}(x) = \int \frac{d^3p}{(2\pi)^3} \frac{1}{\sqrt{2mE(\textbf{p})}}\sum_{h} \bigg[c_{h}(\textbf{p})\psi^{\mathtt{P}}_{h}(p^{\mu})e^{-ip_{\mu}x^{\mu}} + d^{\dagger}_{h}(\textbf{p})\psi^{\mathtt{A}}_{h}(p^{\mu})e^{ip_{\mu}x^{\mu}}\bigg], 
\end{eqnarray}
and the associated dual given by
\begin{eqnarray}\label{campoquanticodual}
\bar{\mathfrak{F}}(x) = \int \frac{d^3p}{(2\pi)^3} \frac{1}{\sqrt{2mE(\textbf{p})}}\sum_{h} \bigg[c^{\dag}_{h}(\textbf{p})\bar{\psi}^{\mathtt{P}}_{h}(p^{\mu})e^{ip_{\mu}x^{\mu}} + d_{h}(\textbf{p})\bar{\psi}^{\mathtt{A}}_{h}(p^{\mu})e^{-ip_{\mu}x^{\mu}}\bigg]. 
\end{eqnarray} The creation and annihilation operators shall obey the usual fermionic relations
\begin{eqnarray}
\lbrace c_{h}(\textbf{p}),c^{\dag}_{h^{\prime}}(\textbf{p}^{\prime})  \rbrace = (2\pi)^{3} \delta^3(\textbf{p}-\textbf{p}^{\prime})\delta_{hh^{\prime}}, 
\\
\lbrace d_{h}(\textbf{p}),d^{\dag}_{h^{\prime}}(\textbf{p}^{\prime})  \rbrace = (2\pi)^{3} \delta^3(\textbf{p}-\textbf{p}^{\prime})\delta_{hh^{\prime}},
\\
\lbrace c_{h}(\textbf{p}),d_{h^{\prime}}(\textbf{p}^{\prime})\rbrace =  \lbrace c^{\dag}_{h}(\textbf{p}),d^{\dag}_{h^{\prime}}(\textbf{p}^{\prime})  \rbrace=0.
\end{eqnarray} Taking advantage of the previous clue concerning the Dirac dynamics for this subclass of spinors, it is possible to find the usual conjugate momentum field as $\pi(x) = i\psi^{\dag}$ and the equal time field-momentum quantum correlator reads 
\begin{eqnarray}\label{correlatorlocalL2}
\{\mathfrak{F}(\vec{x},t), i\mathfrak{F}^{\dag}(\vec{x}\;^{\prime},t)\} = i\delta^3(\vec{x}-\vec{x}\;^{\prime})\I.
\end{eqnarray} 
evincing a local field. 

A clear characteristic appearing in the Weinberg formulation is that the Dirac dynamics satisfied by the expansion coefficients of the quantum fermionic field may be understood as a (full) Lorentz-invariant record of bi-spinors belonging to the representation space $(1/2,0)\oplus (0,1/2)$ \cite{weinberg1}. As previously mentioned, there is a direct relation between parity operator acting upon spinors at the classical level and the Dirac dynamics \cite{speranca}, i. e., $P=m^{-1}\gamma^\mu p_{\mu}$. By its turn, a relation of eigenspinor with respect to parity is indeed valid if the one particle state resulting from the quantum field (acting upon the vacuum state) has not degeneracy beyond the one coming from the spin \cite{weinberg1, estwig}. We refer to these quantum states as standard\footnote{An unusual case, in the sense just described, is discussed in Ref. \cite{elkostates}.}. As we see, a fermionic quantum theory respecting full Lorentz symmetries and ensuring locality is only possible by means of spinors belonging to $L_2$ as expansion coefficients. For a spin $1/2$ quantum field theory, describing particles without further degeneracy, the usual physical requirements may be replaced, then, by the demand of expansion coefficients belonging to $L_2$.    

\subsection{An explicit example for non-locality}

Let us analyse the resulting physical situation when the spinors at hand do not belong to the subclass $L_2$, that is for which $\alpha\neq\beta$. In order to provide a concrete case and fix ideas we shall develop the case $e^{i\alpha}=ia$ and $e^{i\beta}=b$ (with real non-null $a$ and $b$ constants), leading, then, to a Lounesto Type-3 case. The spinors read
\begin{eqnarray}
\psi^{\mathtt{P}}_{(+,+)}(p^{\mu}) = \sqrt{m}\left(\begin{array}{c}
ia\mathfrak{B}_{+}\cos(\theta/2)e^{-i\phi/2} \\ 
ia\mathfrak{B}_{+}\sin(\theta/2)e^{i\phi/2} \\ 
b\mathfrak{B}_{-}\cos(\theta/2)e^{-i\phi/2} \\ 
b\mathfrak{B}_{-}\sin(\theta/2)e^{i\phi/2}
\end{array}\right), \; \psi^{\mathtt{P}}_{(-,-)}(p^{\mu}) = \sqrt{m}\left(\begin{array}{c}
-ia\mathfrak{B}_{-}\sin(\theta/2)e^{-i\phi/2} \\ 
ia\mathfrak{B}_{-}\cos(\theta/2)e^{i\phi/2} \\ 
-b\mathfrak{B}_{+}\sin(\theta/2)e^{-i\phi/2} \\ 
b\mathfrak{B}_{+}\cos(\theta/2)e^{i\phi/2}
\end{array}\right).
\end{eqnarray} 
Under the acting of the Dirac operator these spinors furnish \begin{eqnarray}\label{dirac1tipo3}
(\gamma_{\mu}p^{\mu}-m)\psi^{\mathtt{P}}_{(+,+)}(p^{\mu})= m\sqrt{m}\left(\begin{array}{c}
ia(-1+b/ia)\mathfrak{B}_{+}\cos(\theta/2)e^{-i\phi/2} \\ 
ia(-1+b/ia)\mathfrak{B}_{+}\sin(\theta/2)e^{i\phi/2} \\ 
b(-1+ia/b)\mathfrak{B}_{-}\cos(\theta/2)e^{-i\phi/2} \\ 
b(-1+ia/b)\mathfrak{B}_{-}\sin(\theta/2)e^{i\phi/2}
\end{array}\right), 
\end{eqnarray}
and
\begin{eqnarray}\label{dirac2tipo3}
(\gamma_{\mu}p^{\mu}-m)\psi^{\mathtt{P}}_{(-,-)}(p^{\mu})= m\sqrt{m}\left(\begin{array}{c}
-ia(-1+b/ia)\mathfrak{B}_{-}\sin(\theta/2)e^{-i\phi/2} \\ 
ia(-1+b/ia)\mathfrak{B}_{-}\cos(\theta/2)e^{i\phi/2} \\ 
-b(-1+ia/b)\mathfrak{B}_{+}\sin(\theta/2)e^{-i\phi/2} \\ 
b(-1+ia/b)\mathfrak{B}_{+}\cos(\theta/2)e^{i\phi/2}
\end{array}\right).
\end{eqnarray} The condition necessary to a Dirac dynamics is therefore never reached. 

The spin sums may be straightforwardly adequate from (\ref{spinsumparticula}) and (\ref{spinsumanti}). These spin sums may also be recast in a Lorentz invariant form. According to our previous results, however, a quantum field may also be defined in a similar fashion of (\ref{campoquantico}) and (\ref{campoquanticodual}), this time of Type-3 spinors as expansion coefficients. The locality inspection, however, brings an peculiar element. Since the spinors do not obey Dirac dynamics at classical level, one is not able to write a Dirac-like Lagrangian and extract the conjugate momentum from it. As it may be readily verified, the Klein-Gordon dynamics is, of course, obeyed. Therefore, the only clue one can follow is starting writing a Klein-Gordon spinorial Lagrangian. The conjugated-momentum, thus, is given by
\begin{eqnarray}
\pi(x) = \frac{\partial \bar{\mathfrak{F}}(x)}{\partial t},
\end{eqnarray} and we arrive at the following equal time quantum correlator
\begin{eqnarray}\label{correlator13-1}
\Bigg\{\mathfrak{F}(\vec{x},t), \frac{\partial \bar{\mathfrak{F}}(\vec{x}\;^{\prime},t)}{\partial t}\Bigg\} = i\int \frac{d^3p}{(2\pi)^3}e^{i\vec{p}\cdot(\vec{x}-\vec{x}\;^{\prime})}\I_{(a,b)}+i\int \frac{d^3p}{(2\pi)^3}(\vec{\gamma}\cdot\vec{p})\;e^{i\vec{p}(\vec{x}-\vec{x}\;^{\prime})}.
\end{eqnarray} The second term on the right-hand side of Eq.\eqref{correlator13-1} stands for the delta distribution and hence 
\begin{eqnarray}\label{correlator13-2}
\Bigg\{\mathfrak{F}(\vec{x},t), \frac{\partial \bar{\mathfrak{F}}(\vec{x}\;^{\prime},t)}{\partial t}\Bigg\} =i\delta^3(\vec{x}-\vec{x}\;^{\prime})\I_{(\alpha,\beta)} + i\int \frac{d^3p}{(2\pi)^3}(\vec{\gamma}\cdot\vec{p})\;e^{i\vec{p}\cdot(\vec{x}-\vec{x}\;^{\prime})},
\end{eqnarray} and the non-local aspect of this field is then evident. Note that this result comes from considering the appropriate lagrangian for the case at hands. An eventual connection with the usual case $(\alpha=\beta)$ must be performed previously, leading then to the Dirac lagrangian, and consequently to the same result expressed in Eq. \eqref{correlatorlocalL2}.

We are not advocating that these fields have not physical relevance, but only pointing out the fact that usual physical requirements for a given fermionic quantum field cannot be accomplished for regular spinors other than are not in $L_2$, with a possible exception discussed in the next paragraph. As the field defined upon Dirac spinors belonging to $L_2$, the fields constructed taking into account Type-1 and Type-3 spinors are of spin one half. Both satisfying the Klein-Gordon equation. The expansion coefficients of the former case stand for a complete set of eigenspinors of the parity operator, while for the latter case the expansion coefficients are not related with any discrete symmetry.
 
Perhaps we could end this section by calling attention to the fact that even for the cases of quantum fields constructed upon expansion coefficients not belonging to $L_2$, there are possibilities to be explored in order to achieve a well behaved physical scenario. The point is that all we assert about locality depends explicitly on the used spinorial dual. Possibilities coming from the Clifford algebra, however, has opened the possibility of different duals \cite{rog1,rog2}. This possibility was first raised and gained the status of a theory (the dual theory) in Refs. \cite{aaca,mdobook}. The use of this crevice to circumvent the no-go theorem was performed in Ref. \cite{nogo}. By the same token, recently new fermionic fields, with well behaved quantum counterparts have been found making use of different duals \cite{novonovo}. The output of the dual theory is to find a well defined fermionic dual in order to make the field local and the fermionic theory covariant under Lorentz symmetries. If this procedure is possible for the cases at hand is unclear. We would like, however, to finalize this section by remarking that, if such a dual is possible here, then one can assert that the fermionic one particle states, whatever they describe, must present degeneracy beyond the spin. The reason is the following: as well known, one particle states, say $\Psi$, are defined as eingenstates of the momentum, spin projection and Hamiltonian operators. The introduction of parity\footnote{The parity operator is here denoted in bold face to be distinguish from the one acting upon classical spinors.}, ${\bf P}$, in the quantum Poincar\`e algebra preserves the eigenvalues of the state ${\bf P}\Psi$ with respect to the same previous operators. Therefore, in the absence of additional degeneracy one would have ${\bf P}\Psi\simeq \Psi$. But this clearly contradicts the fact that expansion coefficients in the present case are not annihilated by the Dirac operator. Thus the only possibility left is that the states of a particle show degeneracy beyond spin.

\section{Concluding Remarks and Outlooks}\label{remarks}

We have shown that, from among several possibilities, a fermionic quantum field endowed with Lorentz symmetries, local, and respecting the Dirac dynamics is only possible (for usual fermionic dual) for a subclass, here called $L_2$, of Type-2 spinors according to Lounesto. Also we have explicit shown an example of a non-local fermionic theory just by constructing an quantum field with expansion coefficients not belonging to $L_2$. Showing a precisely consistency of the Weinberg's formalism \cite{weinberg1}, the Dirac field describes fermions under the following two basic requirements: if parity covariance is ensured and by imposing locality in a quantum field theoretic framework. Within this context, only spinors belonging to $L_2$ should have the epithet of Dirac spinors.   

These results may be seen as a link between the quantum fermionic field no-go theorem and the Lounesto classification of classical spinor fields. This link points out what classical spinors lead to a non-pathological quantum theory. For the other cases, it is our current understanding that if there is a chance for a well behaved quantum theory even using spinors not belonging to $L_2$ then this chance must makes use of different duals. Otherwise one would have, eventually, to embrace a non-local theory. In this case, it would be interesting to wonder about the mass generation analogues of the Higgs mechanism. We remark by passing that additional trial phases may be used in the bi-spinor composing, leading essentially to the same results.

In the light of our remarks along the text, theories coming from a sector whose spinors are not in $L_2$, but with a new dual leading to a well behaved final format, shall describe one particle states with degeneracy beyond the spin. In a Universe whose known constituents perform four percent, these new possibilities certainly may be faced as intriguing. Obviously, by using duals different from the standard one, the Lounesto classification must be revisited \cite{beyondlounesto}. This is a relevant point to further explore: what subclass of spinors (if any) in the new classification would lead to a well behaved fermionic quantum field.

\section{Acknowledgements}
The authors express their gratitude to Cheng-Yang Lee for insightful questions and ensuing discussions during the manuscript writing stage. JMHS thanks to CNPq grant No. 303561/2018-1 for partial support.

\end{document}